\documentclass[twocolumn,secnumarabic,amssymb, amsmath, nofootinbib,tightenlines,
nobibnotes, aps, prl]{revtex4}
\usepackage{amsmath}%
\usepackage{longtable}%
\usepackage{bm}%
\expandafter\ifx\csname package@font\endcsname\relax\else
 \expandafter\expandafter
 \expandafter\usepackage
 \expandafter\expandafter
 \expandafter{\csname package@font\endcsname}%
\fi

\usepackage{graphicx}
\usepackage{graphicx}

\begin{document}

\title{ Multiscaling  in Strong Turbulence Driven by a Random Force. }
 \newcommand {\BU}{Department of Mechanical Engineering, Boston University, Boston, Massachusetts, 02215,USA}

\author{Victor Yakhot}
\affiliation{\BU}
\author{ Diego Donzis }
 \affiliation{Department of Aerospace Engineering, Texas A\&M University, College Station, Texas 77843,USA}


\email[]{vy@bu.edu}

\date{\today}

\begin{abstract}

 \noindent  Turbulence problem is often considered as ``the last unsolved problem of classical physics''. It is due to strong interaction between velocity and/or  velocity gradient fluctuations,  
 a high Reynolds number flow is  a fascinating  mixture of purely random, close to Gaussian,  fields and coherent structures where substantial fraction of kinetic energy is dissipated into heat. To evaluate  intensity of fluctuations, one usually studies  different  moments of velocity increments  and/or  dissipation rate,   characterized by scaling exponents  $\zeta_{n}$ and $d_{n}$, respectively.  In high Reynolds number flows,  the moments of different orders with $n\neq m$  cannot be simply related to each other, which is the signature  of  anomalous scaling,  making this   problem ``the last unsolvable''.  No perturbative treatment can lead to quantitative description of this feature.
 In this work the  expressions  for the moments  of  dissipation rate $e_{n}=\overline{{\cal E}^{n}}\propto Re^{d_{n}}$   and those of velocity derivatives  $M_{2n}=\overline{(\partial_{x}u_{x})^{2n}}\propto \frac{v_{o}^{2n}}{L^{2n}}Re^{\rho_{2n}}$  are  derived for an infinite  fluid  stirred by a white-in-time Gaussian random  force supported in the vicinity of the wave number $k_{f}\approx \frac{2\pi}{L}=O(1)$, where $v_{0}$ and  $L$ are characteristic velocity and integral scale, respectively.  A novel aspect of this work is that unlike previous efforts which aimed at seeking solutions around the infinite Reynolds number limit, we  concentrate on  the  vicinity of transitional Reynolds numbers $Re^{tr}$ of the first emergence of anomalous scaling out of Low-Re Gaussian background.  The obtained closed expressions    for anomalous scaling exponents $d_{n}$ and $\rho_{n}$   agree well with   available in  literature experimental  and numerical data  and,  when $n\gg 1$,   $d_{n}\approx  0.3n \ln(n)$.

 \end{abstract}

\maketitle
PACS numbers 47.27\\

\noindent {\bf   Introduction.} \\

\noindent {2. \it ``Reynolds numbers'' . Definitions.}  If  a velocity gradient field  obeys   Gaussian statistics with the Re-dependent variance, 
$\overline{(\partial_{x}v_{x})^{2}}\propto \frac{v_{0}^{2}}{L^{2}}Re^{\rho_{2}}$, its moments 
$\overline{(\partial_{x}v_{x})^{2n}}=(2n-1)!!( \overline{(\partial_{x}v_{x})^{2}})^{n}
\propto \frac{v_{o}^{2n}}{L^{2n}}Re^{n\rho_{2}}$.  
Here, $v_{0}=O(1.)$ and $L=2\pi/k_{0}=O(1)$ are  the  single-point large-scale properties of the flow. 
The above-mentioned`` normal scaling''  is not the only possibility.  Indeed,  high-Reynolds-number flows are characterized by ``anomalous'' scaling exponents reflecting the formation  of coherent structures. In this case $\overline{(\partial_{x}v_{x})^{2n}}\propto \frac{v_{0}^{2n}}{L^{2n}}Re^{\rho_{2n}}$ where $\rho_{2n}\neq n\rho_{2}$.  
In principle, depending on the dynamics, the transitional  Reynolds numbers of  different moments  $n$ of velocity field may be different and with increase of the Reynolds number above critical one ($Re_{n}^{tr}$), the flow may become a superposition of Gaussian and ``anomalous'' patches obeying different laws. 
This  fact makes the problem of anomalous scaling or intermittency so difficult. 

\noindent In general, we have to   define an infinite number of large-scale mean velocities  $\hat{v}(n)=L\overline{(\partial_{x}v_{x})^{n}}^{\frac{1}{n}}$  and corresponding  Reynolds numbers:

 \begin{equation}
\hat{R}e_{n}= \hat{v}(n)L/\nu\equiv \frac{L^{2}\overline{(\partial_{x}v_{x})^{n}}^{\frac{1}{n}}}{\nu}\propto Re^{\frac{\rho_{n}}{n}+1}
\end{equation}

\noindent where $\rho_{n}$ is a scaling  exponent of the $n^{th}$ -order moment of velocity derivative, i.e. $\overline{(\partial_{x}v_{x})^{n}}\propto Re^{\rho_{n}}$.  As follows from (1),


\begin{equation}
Re\approx \hat{R}e_{n}^{\frac{n}{\rho_{n}+n}}
\end{equation}

\noindent and,  the widely used in hydrodynamics large-scale dimensionless parameter, the  ``Reynolds number'', is : $Re\equiv \hat{R}e_{2}=\frac{(\partial_{x}v_{x})_{rms}L^{2}}{\nu}$. 
One can  introduce an infinite number of  Reynolds numbers based on ``Taylor length scales'':

\begin{eqnarray}
\hat{R}_{\lambda,n}= \sqrt{\frac{15L^{4}}{\overline{3{\cal E}\nu}}}\hat{v}(2n)=\nonumber \\
\sqrt{\frac{15v_{0}^{3}}{\overline{3L}{\cal E}}}\sqrt{Re}\times \frac{L^{2}}{v_{0}^{2}}\overline{(\partial_{x}v_{x})^{2n}}^{\frac{1}{n}}\approx Re^{\frac{\rho_{2n}}{n}+\frac{1}{2}}
\end{eqnarray}

\noindent  where   dimensionless combination $\frac{{\cal E}L}{v_{0}^{3}}=O(1)$. It  clear that the proportionality coefficient in (3) is $n$-independent. This fact will be essential for what follows.   Multiplying and dividing right side of (3) by viscosity $\nu$  and taking into account that in the system under consideration ${\cal E}=P=const=O(1)$ (see below),  one obtains $Re^{\frac{\rho_{2n}}{n}}\approx Re^{\frac{d_{n}}{n}+1}$ giving 

$$\rho_{2n}=d_{n}+n$$              
  
If the flow is generated by a Gaussian random force (see below), in the limit $Re\rightarrow 0$,  the velocity field is random and at least  the first few low-order moments  obey close-to-Gaussian statistics.  The question arises:  given the random low-Reynolds number flow,  how do we define  chaotic ``strong'' turbulence dominating the large-Re dynamics?  This problem was addressed in  Ref.[1] where we discussed emergence of intermittency out of ``normal''  Gaussian background with increase of $Re$.  A similar question  has been studied   by Kuz'min and Patashinsky  [2] in terms of small-scale phase organization   in a low Reynolds number flow. 

\noindent  
It has been demonstrated both theoretically and numerically  in Ref.[1] that transitional Reynolds  number  $\hat{R}e_{2}^{tr}\equiv Re^{tr}$ from ``normal'' to ``anomalous''   scaling of normalized velocity derivatives 
$M_{2n}=\overline{(\partial_{x}v_{x})^{2n}}\propto (\frac{v_{o}}{L})^{2n}Re^{\rho_{2n}}$  and those  of dissipation rates  $e_{n}=\overline{{\cal E}^{n}}\propto Re^{d_{n}}$
are  $n$-dependent,  monotonically decreasing with increase of the moment order $n$ [1],[3] ( also see Fig.1).  It was further shown that   $Re^{tr}\approx 100$ ($R^{tr}_{\lambda}\approx 9.$) for $2n\approx 4$.  {\bf However, expressed in terms of $\hat{R}_{\lambda,n}$, the observed transition points   $\hat{R}^{tr}_{\lambda,n}$ were  $n$-independent with 
  $\hat{R}^{tr}_{\lambda,n}\approx 9.0$.  In this paper, using this effect 
   as a dynamic constraint, we generalize the theory developed in [1] to calculate anomalous scaling exponents $\rho_{n}$ and $d_{n}$.}

\noindent   1. {\it Background.} In 1965, in his classic textbook [4],  R.P. Feynman proclaimed   ``turbulence as the most important unsolved  problem of classical physics''.  His point was that   by 1965 we knew more about weak and strong interactions of elementary particles  than about  flow of water out of   faucets  in our bathtubs. 
Feynman  did not elaborate  on his  statement and we  can only guess that  he had in mind a  derivation of chaotic,  high-Reynolds number   solutions  directly from  deterministic and well-known Navier-Stokes equations for incompressible fluids.  
By that time it became clear (for a comprehensive review see Refs. [5])  that, due to the  lack of a small coupling constant,  the problem of  derivation of the  energy spectra  and other correlation functions using renormalized perturbation expansions was  at least as hard as  that  in the  theory of strong interactions.  
At  approximately the same time  in an extraordinary  1969 paper, Polyakov,   developing  his  theory of  strong interactions  of hadrons [6],   
accounted  for  the  multi-particle  (high-order) contributions to the diagrammatic expansions of   Green functions,   leading to  the first calculation of anomalous scaling exponents ({\bf multiscaling}).
Unlike scaling solutions in the theory of critical point  or Kolmogorov's similarity relations for the low-order moments of velocity increments,   in multiscaling  the  moments of  different orders are characterized by scaling functions of different dimensionalities. 
To illustrate the physics behind this novel concept, Polyakov constructed a  qualitative ``cascade model''  yielding  the calculated    scaling exponents.  
Since 1941 various cascade models  leading to Kolmogorov's energy spectrum, have been well-known  (for  a very detailed review and discussions see Ref. [5]-[6]).  Starting from 1962 Kolmogorov's attempts to derive anomalous scaling exponents, numerous ``cascade models'' have been developed to obtain exponents for high-order structure functions and moments of dissipation
Regretfully, these  ideas,  applied  to strong  turbulence,  failed.   Anyhow, Feynman's  assessment   of the ``turbulence problem''  has been repeated countless  times as an inspiration   for  various attacks  on this problem. \\
\noindent Twenty seven years later, in a {\bf Science}  paper,  M.Nelkin,  reviewing  the status  of turbulence research,  asked a crucial question:  {\bf ``In What Sense Turbulence in Unsolved Problem?'' } [7].   
Usually,  fluid  turbulence is  a result of instability of a low-Reynolds number laminar flow which is a solution to  the Navier-Stokes equations in the limit $Re=VL/\nu\rightarrow 0$, where $V$ and $L$ are free-stream velocity and characteristic flow length-scale, respectively. 
Since the number of possible laminar flows is, for all practical purpose infinite, one can hardly expect  a  large-scale universality of both  mechanisms  of  turbulence production and of magnitudes of critical  (transitional) Reynolds number.   Still, according to  Kolmogorov's   theory of 1941,   one can expect some  universality  on the scale $r/L \ll 1$ and it is possible that  in this interval  the size of the system disappears from the dynamics: the  energy spectrum then is given by the scaling law:

$$E(k)=\overline{{\cal E}}^{\frac{2}{3}}k^{-\frac{5}{3}}f(k\eta)$$

\noindent where $\eta$ is the dissipation scale (see below). 
The scaling function $f(x)$ is assumed universal. This relation can be rewritten in  physical space
\begin{equation}
S_{2}(r)=\overline{(u(x+r)-u(x))^{2}}=(\overline{{\cal E}}r)^{\frac{2}{3}}f(\frac{r}{\eta})
\end{equation}

\noindent  At the length-scale $r\approx \eta$  viscous effects become important and in the interval $r\ll \eta$, velocity field is  an {\bf analytic function}.  It was further assumed that in the inertial range  $L\gg r \gg \eta$ the viscous effects are negligibly weak   and the scaling function $f(x)\rightarrow C=const$. If this is so,  one obtains the so called Kolmogorov's spectrum $E(k)\propto C{\cal E}^{\frac{2}{3}}k^{-\frac{5}{3}}$ and $S_{2}\propto r^{\frac{2}{3}}$. These considerations, leading to the above  ``$5/3$ energy spectrum'',  are often called in the literature: ``Kolmogorov's K41 theory''.  
Nelkin stressed that, while there was no solid theoretical derivation of these results, they were quite close to the available at the time experimental data.  \\

\noindent  The relation (3) leads to an important conclusion: consider $r/\eta\rightarrow 0$.   In this limit the velocity field is an analytic function, so that $S_{2}\approx \overline{(\frac{\partial_{x}u_{x}}{\partial x})^{2}}r^{2}=\overline{{\cal E}}r^{2}/(15\nu)$,  and we can safely set $r\rightarrow\eta$. Taking into account that, by the energy conservation,  $\overline{{\cal E}}L/V^{3}=O(1)$,  we, matching analytic and inertial (singular) ranges,  obtain   $(\frac{\eta}{L})^{\frac{2}{3}}\approx Re(\frac{\eta}{L})^{2} $, which is nothing but Kolmogorov's relation $\eta=L Re^{-\frac{3}{4}}$ where the large-scale Reynolds number $Re=VL/\nu$.
Small correction to Kolmogorov's spectrum   ( $E(k)\propto k^{-0.71}$ ) has been reported  (see Ref. [8],[9]).   

\noindent The situation is totally unsatisfactory when K41 is  used to predict high-order moments of velocity increments and dissipation rate.  Indeed,  K41 gives 

$$\overline{{\cal E}^{n}}=\nu^{n} \eta^{\frac{4n}{3}}\approx  Re^{0}=const$$

\noindent This result contradicted all known experimental data which  gave strong indications of  Kolmogorov's theory breakdown  at small scales $r \rightarrow \eta$ [5],[7]- [9].   Moreover, the scaling solution 

$$S_{n}=r^{-\gamma}\int (\delta_{r}u)^{n}{\cal P}(\frac{\delta_{r}u}{r^{\gamma}})d(\delta_{\eta}u)\propto r^{n\gamma}$$

\noindent  where $r^{-\gamma}P(x)$ is a scaling function for the normalized probability density  (PDF), of velocity increments,   
failed to describe the experimentally observed multiscaling,  defined by the relation $S_{n}\propto r^{\zeta_{n}}$  with  the exponents $\zeta_{n}/n\neq  \zeta_{m}/m$ for $n\neq m$.   
Nelkin stressed the  inability of scaling theory  to describe  the moments  of dissipation rate $e_{n}=\overline{{\cal E}^{n}}\propto Re^{d_{n}}$  as the  main deficiency of exiting  turbulence theory.  This was one  of his  reasons to deem turbulence problem ``the last...''.\\

\noindent The failure of the formula $\eta\propto LRe^{-\frac{3}{4}}$ has been relatively recently theoretically  predicted and  detected in  very accurate numerical simulations [9], where it was shown that the more accurate dependence was $\eta\equiv \eta_{2}=LRe^{-0.78}$. While this deviation  from  Kolmogorov's $\eta_{2}\approx LRe^{-0.75}$ can be regarded as small, the experimentally observed ``dissipation scales''  $ \eta_{n}\approx LRe^{-\theta_{n}}$,   with with $n>2$,  
very  strongly deviated   from Kolmogorov's value $\theta_{n}=3/4$. The length-scales  $\eta_{n}$ are defined as follows: $S_{n}=\overline{(u(x+r)-u(x))^{n}}\propto r^{n}$ when $r/\eta_{n}\ll 1$.
The explanation of this phenomenon is hidden deep in the dynamics of turbulence and turbulence transition. If, following Landau/Lifshitz, one assumes [10]  that turbulence transition  happens locally when $Re_{r}=(\delta_{r}u)r/\nu\geq  1$, one concludes that the flow is a superposition of turbulent  ($Re_{r}>1$) and laminar   ($Re_{r}<1$),  patches.  Therefore, locally,  the dissipation (viscous) scale, given by $Re_{\eta}\approx 1$ or 
\begin{equation}
\eta\approx  \nu/\delta_{\eta}u
\end{equation}

\noindent is a  strongly fluctuating parameter depending on the local magnitude of $\delta_{\eta}u$. That is where Kolmogorov's theory breaks down giving rise to huge anomalous fluctuations of all spatial derivatives, including dissipation rate ${\cal E}$.  The relation (5) gives the low bound on the ``dissipation scale''.  It is clear that the   largest  velocity fluctuations are $u\approx u_{rms}$. Indeed  due to the Gaussian statistics of large-scale fluctuations, the probability of  an event $u>u_{rms}$ is very small.  Substituting this estimate into (4) gives  in the high-Re limit:

$$\eta_{min}/L\approx 1/Re<<Re^{-\frac{3}{4}}$$

\noindent In the limit $Re\rightarrow\infty$, huge    fluctuations  of dissipation scale make renormalized perturbations  expansions problematic.\\

\noindent In this paper we   develop a quantitative theory   for the  moments of velocity derivative  in  a particular problem of a random-force-driven Navier-Stokes turbulence.\\

 \noindent  {\bf  The model}.  Fluid flow can be described by the Navier-Stokes equations subject to boundary and initial conditions (the density is taken $\rho=1$ without loss of generality): 

\begin{equation}\partial_{t}{\bf u}+{\bf u\cdot\nabla u}=-\nabla p +\nu\nabla^{2}{\bf u} + {\bf f} \end{equation}

\noindent   $\nabla\cdot {\bf u}=0$.   A random Gaussian noise ${\bf f} $ is  defined by the correlation function [10],[11]:

 \begin{equation}\overline{f_{i}({\bf  k},\omega)f_{j}({\bf k'},\omega')}= (2\pi)^{d+1}D_{0}(k)P_{ij}({\bf k})\delta({\hat{k}+\hat{k}'})\end{equation}

\noindent  where the four-vector $\hat{k}=({\bf k},\omega )$ and projection operator is: $P_{ij}({\bf k})=\delta_{ij}-\frac{k_{i}k_{j}}{k^{2}}$.    In  equilibrium fluid   the  thermal fluctuations, responsible for Brownian motion   are generated by the forcing (6) with $D_{0}(k)= \frac{k_{B}T \nu}{\rho}k^{2}$ [10],[11]. In channel flows or boundary layers 
with  rough walls, the amplitude $D_{0}$ is of the order of the rms magnitude of the roughness element [2]. 
Here we are interested in the case $D_{0}(k)=const \neq 0$ only in the interval close to    $k\approx 2\pi/L$, discussed by Forster, Nelson and Stephen  [11].
The energy balance, written here for the case of isotropic and homogeneous flow, following from (5) -(6) imposes the energy conservation constraint:
${\cal P}=\overline{{\bf u}\cdot {\bf f}}={\cal E}=\frac{\nu}{2}\overline{(\frac{\partial u_{i}}{\partial x_{j}}+\frac{\partial u_{j}}{\partial x_{i}})^{2}}= \nu\overline{(\frac{\partial u_{i}}{\partial x_{j}})^{2}}= O(1)$, 
where  energy production rate  ${\cal P}=O(1)$  is an external parameter independent upon Reynolds number.   
The random-force-driven NS equation can be written in the Fourier space:
\begin{widetext}
\begin{equation}
 u_{l}({\bf k},\omega)=G^{0}f_{l}({\bf k,\omega})-
\frac{i}{2}G^{0}{\cal P}_{lmn}\int u_{m}({\bf q},\Omega)u_{n}({\bf k-q},\omega-\Omega)d{\bf Q}d\Omega
\end{equation}
\end{widetext}
\noindent where $G^{0}=(-i\omega+\nu k^{2})^{-1}$,  ${\cal P}_{lmn}({\bf k})=k_{n}P_{lm}({\bf k})+k_{m}P_{ln}({\bf k})$  and,  introducing the zero-order solution ${\bf u}_{0}=G^{0}{\bf f} \propto \sqrt{D_{0}}$, so that ${\bf u}=G^{0}{\bf f}+{\bf v}$,  one derives the equation for perturbation ${\bf v}$:

\begin{widetext}
\begin{eqnarray}
v_{l}(\hat{k})=-\frac{i}{2}G^{0}(\hat{k}){\cal P}_{lmn}({\bf k})\int v_{m}(\hat{q})v_{n}(\hat{k}-\hat{q})d\hat{q}\nonumber\\
-\frac{i}{2}G^{0}(\hat{k}){\cal P}_{lmn}({\bf k})\int  [v_{m}(\hat{q})G^{0}(\hat{k}-\hat{q})f_{n}(\hat{k}-\hat{q})+ G^{0}(\hat{q})f_{m}(\hat{q})v_{n}(\hat{k}-\hat{q})]d\hat{q} \nonumber \\
-\frac{i}{2}G^{0}(\hat{k}){\cal P}_{lmn}({\bf k})\int G^{0}(\hat{q})f_{m}(\hat{q})G^{0}(\hat{k}-\hat{q})f_{n}(\hat{k}-\hat{q})d\hat{q}
\end{eqnarray}
\end{widetext}

\noindent  It is clear from (8) that the correction to the $O(\sqrt{D_{0}})$ zero-order Gaussian solution is driven 
by  the ``effective forcing''

\begin{eqnarray}
F_{l,2}=-\frac{i}{2}G^{0}(\hat{k}){\cal P}_{lmn}({\bf k}) \times\nonumber\\ \int G^{0}(\hat{q})f_{m}(\hat{q})G^{0}(\hat{k}-\hat{q})f_{n}(\hat{k}-\hat{q})d\hat{q}=O(D_{0})\nonumber
\end{eqnarray}

\noindent which is small in the limit $D_{0}\rightarrow 0$.   We can define $v_{\l}=v_{1,l}+G^{0}F_{l,2}$ etc and generate renormalized expansion in powers of dimensionless  ``coupling constant'' $\lambda\propto D_{0}$ resembling that formulated in a classic work by  Wyld (see Ref. [12]).  However,  this expansion, is   haunted by divergences [5] 
making its analysis  extremely hard if not impossible.  In this work, we will try to avoid perturbative treatments of  equation of motion (7)-(8).

According to  Landau and Lifshitz [10], if $\delta_{\eta}v\approx v(x+\eta)-v(x)$,  the dissipation scale $\eta$ is defined by a condition $Re_{\eta}\approx1=\eta (\delta_{\eta}v )/\nu$, so that $\eta=\nu/(\delta_{\eta}v)$. 
Thus, the simple algebra gives: $\partial_{x}v_{x}\approx (\delta_{\eta}v_{x})^{2}/\nu$. Therefore

$$(\frac{L}{v_{0}})^{2n}\overline{(\frac{\partial v_{x}}{\partial x})^{2n}}=Re^{2n}(\frac{\eta_{4n}}{L})^{\zeta_{4n}}=Re^{\rho_{2n}}$$

\noindent  where $Re=v_{0}L/\nu$ and  below we set the large-scale properties $v_{0}\equiv v_{rms}=1$ and $L=1$, so that: 

$$
\frac{L^{n}}{v_{0}^{3n}}\overline{{\cal E}^{n}}=Re^{d_{n}}=Re^{n}
\frac{\overline{(\delta_{\eta}v)^{4n}}}{v_{0}^{4n}}
=Re^{n}S_{4n}(\eta_{4n})=Re^{n}(\frac{\eta_{4n}}{L})^{\zeta_{4n}}
$$

\noindent  where the scaling exponents $d_{n}$ and $\rho_{n}$ are yet to be derived from an  a'priori theory 
as well as the exponents $\zeta_{n}$  entering  the so called ``structure functions''  as: $S_{n}(r)=\overline{(v(x+r)-v(x))^{n}}\propto r^{\zeta_{n}}$.   The dissipation rate ${\cal E}=\nu S_{pq}^{2}$ with $S_{pq}=\frac{\nu}{2}(\partial_{p}v_{q}+\partial_{q}v_{p})$  includes various   derivatives 
$\partial_{p}v_{q}$ with  $p=q$  and $p\neq q$.  Below, based on isotropy, we will use $\overline{{\cal E}}=
15\nu\overline{(\partial_{x}v_{x})^{2}}=v_{0}^{3}/L$  as a normalization factor yielding  dimensionless dissipation rate ${\cal E}=1$. This way we assume that all  contributions  to the moments $\overline{{\cal E}^{n}}$  scale the same way. Thus, in what follows we will be working in the units defined by the large-scale properties of the flow $v_{0}=L=1$ with $Re=1/\nu$.  In the vicinity of transition, 
 when the forcing is  supported  in a  narrow interval in  the wave-vector space,  $(\partial_{x}v_{x})_{rms}\approx (v_{x}(x+L)-v_{x}(x))_{rms}/L\approx v_{rms}=v_{0}$. In a Gaussian case, where $\rho_{2n}=\rho_{2}n$, all Reynolds numbers are of the same order and 
each one, for example $Re_{1}$,  is sufficient for description of a flow. \\


\noindent {\bf  Transition.}   According to theory and numerical simulations of Refs.  [1],[9], [13]-[16] , $Re^{tr}_{2}\approx 100.0$ or $R^{tr}_{\lambda,2} \approx 9.0$ and  
this point we associate with transition to strong  turbulence,  characterized by non-Gaussian statistics of velocity field and by anomalous scaling or ``intermittency'' of increments and velocity derivatives.  {\bf The transition points of high-order moments  with $n>2$, expressed in terms of the standard second-order Reynolds number $Re=v_{rms}L/\nu$,   are  $Re^{tr}<100.0$  ($R^{tr}_{\lambda}<9.0$).
 In  turbulent flows the fluctuations with $\hat{v}(n)=\overline{v^{n}}^{\frac{1}{n}}>v_{rms}$ do exist and one can expect   transitions when {\bf local}  $\hat{Re}_{n}=\hat{v}(n)L/\nu \geq 100.0$ even in  
 low Reynolds number subcritical flows  with   $Re=\hat{Re}_{2}=v_{rms}L/\nu< 100.0$  or $R_{\lambda}\equiv R_{\lambda ,2} <9.0$.  This effect has been clearly demonstrated in numerical simulations of isotropic turbulence [1]  (also, see Fig.1 and Supplemental Material) )  and  in experiments  in  ``noisy''  channel flow with randomly rough walls [3].  }

{\bf To summarize: critical  Reynolds  numbers  $Re^{tr}=(\partial_{x}v_{x})_{rms}L^{2}/\nu$ 
 for the   $n^{th}$-order  moments  of velocity increments  (spatial derivatives)   are  $n$-dependent.  At the same time, expressed in  terms of the conditional Reynolds numbers 
$\hat{Re}_{n}$,  based on  characteristic velocities  $\hat{v}(n)$, the transition occurs at $\hat{R}e_{n}^{tr}\approx 100$ or $\hat{R}_{\lambda,n}\approx 9.0$, independent on the moment order $n$.}

\noindent  
Thus, we  have found   a  new invariant [1]:  at transition points of different moments:

$$\hat{R}^{tr}_{\lambda,n}= R^{tr}_{\lambda,2}\approx 9.0=const$$

\noindent  where $\hat{R}_{\lambda,n}$,  is given by expression (3).     


\noindent  2. {\it Evaluation of $R_{\lambda}\equiv R_{\lambda,2}$.} In the linear (Gaussian) regime  only the modes ${\bf v(k)}$ with $k \approx 2\pi/L$ are excited and   in the vicinity of a transition point we can define the Taylor-scale Reynolds number:
 
\begin{eqnarray}
R_{\lambda}\equiv R_{\lambda,2}=\sqrt{\frac{5}{3{\cal E}\nu}}v_{rms}^{2}=
\frac{v_{rms}L}{\nu}\sqrt{\frac{5v_{rms}^{2}\nu}{3{\cal P} L^{2}}}\nonumber \\
 \approx Re\times \sqrt{\frac{5v_{rms}^{3}}{3 {\cal P}LRe}}\approx  \sqrt{Re/1.2}
\end{eqnarray}

\noindent  Thus,  at a transition point,  where the  theoretically predicted and confirmed in many numerical simulations $R_{\lambda,tr}\approx 8.909$ Refs. [1], [9],[14]-[16] (also see Fig.1) we obtain 
$Re^{tr}\approx 1.2 R^{2}_{\lambda,tr}\approx 120$, close to the one  obtained in numerical simulations  [9]. This estimate is  based on the  following from (5)-(6) assumption of  {\bf a constant, independent on the Reynolds number},  dissipation rate ${\cal E}={\cal P}=const$ and Kolmogorov-like estimate  $u_{rms}^{3}\approx {\cal P}L$ obtained with the Kolmogorov's constant $C_{K}\approx 1.65$  and at the internal scale $2\pi/L\approx 1$.\\

\noindent {\bf  Relation between exponents $\rho_{n}$ and $d_{n}$ when  ${\cal E}=P=const$.} 

\noindent  By the energy balance following equations of motion  (6)-(8),  ${\cal  E}=P=1$, independent on Reynolds number.   In the limit $Re\rightarrow 0$, we have:

$$M_{2n}^{<}=\frac{{\overline{(\partial_{x}v_{x} )^{2n}}}}{\overline{(\partial_{x}v_{x})^{2}}^{n}}=\frac{\overline{{\cal E}^{n}}}{\overline{{\cal E}}^{n}}=\overline{{\cal E}^{n}}=(2n-1)!! $$

\begin{figure}[h]
\includegraphics[width=9cm]{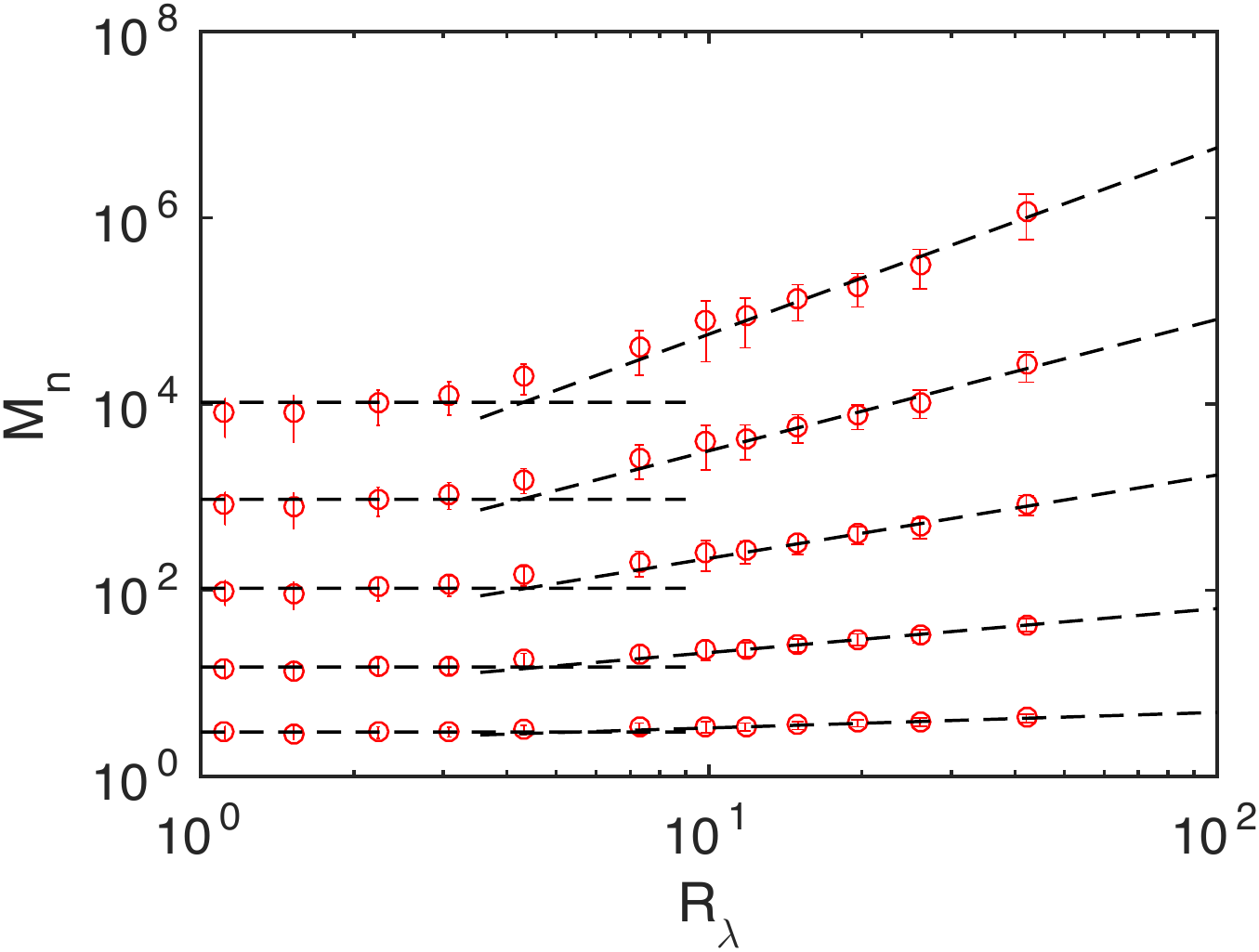}
\caption{ Dimensionless moments $M_{2n}(R_{\lambda})$  from Ref.[1].   Dashed lines: theoretical predictions and numerical simulations of Refs.[1], [9].  Squares are from Ref.[9] and asterisks from our large DNS data base (see Ref. [13] );   Horizontal lines mark levels  of Gaussian moments $M_{2n}=(2n-1)!!$.    Transition points are at the crossings  of  horizontal and dashed lines with transitional $R^{tr}_{\lambda,2}\approx 9.0$  clearly seen on a graph for $M_{2}$. }
\end{figure}

 \noindent  and,   seek  the large-Re solution in the form:

\begin{equation}
M_{2n}^{>}=\overline{(\partial_{x}u_{x})^{2n}}\propto  (\frac{v_{0}}{L})^{2n}Re^{\rho_{2n}}
\end{equation}

\noindent  
 Multiplying   (11) by $\nu^{n}$ gives:
 
 \begin{equation}
 (\frac{L}{v_{0}^{3}})^{n}\overline{{\cal E}^{n}}=Re^{d_{n}}=Re^{\rho_{2n}-n}
 \end{equation}

\noindent leads to the relation between scaling exponents $d_{n}$ and $\rho_{2n}$:

\begin{equation}
\rho_{2n}=d_{n}+n
\end{equation}

\noindent  obtained in Introduction. This  relation,   which is is a consequence of the energy conservation law in the flow driven by white-in-time random force,  was studied in Ref.[9].  {\bf  We would like to stress that 
 the white-in-time property of the stirring force is crucial, which  makes numerical simulations of the moments of dissipation rate  in  the low-Reynolds number flows somewhat difficult.}
 


\noindent  {\bf  Scaling exponents.}  
As follows from the definition (3),  at a transition point of the $n^{th}$ moment,   the large-scale Reynolds number is:

\noindent and 
\begin{equation}
Re^{tr}_{n}=C(\hat{R}^{tr}_{\lambda,n})^\frac{1}{\frac{d_{n}}{n}+\frac{3}{2}}
\end{equation}

\noindent  with   {\bf the $n$-independent}  factor $C$.  
Since $R_{1}^{tr}\approx 100-200$, $d_{1}=0$ (see below)  and $R^{tr}_{\lambda,1}\approx 8.91$, calculated above,  we obtain  an estimate $C\approx  50-100$. 
As follows from (10) and  (13), in the vicinity of transition with $d_{1}=0$:

\begin{equation}
\overline{{\cal E}^{n}}=(2n-1)!!=C^{d_{n}}(R^{tr}_{\lambda,n})^{\frac{nd_{n}}{d_{n}+\frac{3n}{2}}}
\end{equation}

\noindent  the closed equation for the anomalous exponents $d_{n}$.   Using large - scale numerical simulations, it  has recently been shown   that  while the large-scale  transitional Reynolds number $Re_{n}^{tr}$ depends on the moment order $n$, the one based on a Taylor scale $R^{tr} _{\lambda,n}\approx 9.0$ is independent on $n$ [1]. {\bf This  result can be readily  understood in terms of the dynamics of transition to turbulence which is not a statistical feature but a property of each realization 
where $R_{\lambda}>R^{tr}_{\lambda}$. In other words all fluctuations  with ``local'' $Re_{\lambda,n}\geq Re^{tr}\approx 9.0.$ undergo transition to turbulence. This argument, consistent with Landau's theory of transition,  fixes  the amplitude in relations (14)-(15) and  enables  evaluation of the scaling exponents by matching two different flow regimes.}
Taking $\ln 8.91\approx 2.48$  gives:

\begin{eqnarray}
d_{n}=-\frac{1}{2}[n(\frac{2.48}{\ln C}+\frac{3}{2})-\frac{\ln (2n-1)!!}{\ln C}]+ \nonumber \\ 
\sqrt{\frac{1}{4} [n(\frac{2.48}{\ln C}+\frac{3}{2})-\frac{\ln (2n-1)!!}{\ln C}]^{2} +\frac{3}{2}n\frac{\ln (2n-1)!! }{\ln C}}
\end{eqnarray}
%

\begin{table}
\begin{ruledtabular}
\begin{tabular}{ccccccccccc}
\hline
$d_{n}$  & $T1$ & $DNS$ &  $C90$ & $Exp.$  \\
\hline
$ d_{1} $ & $0.00$ & $0.00$ &  $0.00 $ & $0.00$ \\
$d_{2} $& $0.158 $& $0.149$ & $ 0.187$ & $0.152$\\
$ d_{3}$ &$ 0.49$& $0.443$ & $0.46$& $0.4$ \\
 $d_{4} $&$0.94$&$ 0.89 $& $0.80$ &$ 0.73$ \\
 $d_{5}$ &$1.49$& $1.47$ &$ 1.19$  & $1.1$\\
 \end{tabular}
 \end{ruledtabular}
\caption{Comparison of theoretical predictions for  exponents $d_{n}$,  given by relation (16),  with experimental  data, semi-empirical models and numerical simulations: $T1$: Theory  [9];  $DNS$;  direct simulations [9], [13] ;  $MF$:  multi-fractal theory [9];  $\ln[C]=4.5$    {\it Exp.} : $d_{n}$ (expression (16) )   with $n!$ instead of $(2n-1)!!$ for the  Low-Re moments. }
\end{table}

\begin{table}
\begin{ruledtabular}
\begin{tabular}{ccccccccccc}
\hline
 $ \rho_{n} $ & $MF$ &$ C90$& $DNS2$\\
\hline
$\rho_{1} $&  $0.474 $& $0.5$& $0.455$\\
$ \rho_{3}$ &$1.57$& $1.61$ & $1.478$\\
$ \rho_{4}$ & $2.19 $& $2.19$ & $2.05$\\
 $ \rho_{5}$& $2.84$&  $2.86 $& $2.66\pm 0.14$\\
 $ \rho_{7} $& $ 4.20 $& $4.17$ & $3.99\pm 0.65$\\
 \end{tabular}
 \end{ruledtabular}
\caption{Comparison of exponents $\rho_{2n}=d_{n}+n$  with the outcome of numerical  simulations and  semi-empirical models.   $MF$ and $DNS2$:  multi-fractal theory  and numerical simulations [9];  $C90$:   expression (16)  with  the constant $C=90.$ }
\end{table}

\begin{figure}[h]
\includegraphics[height=5cm]{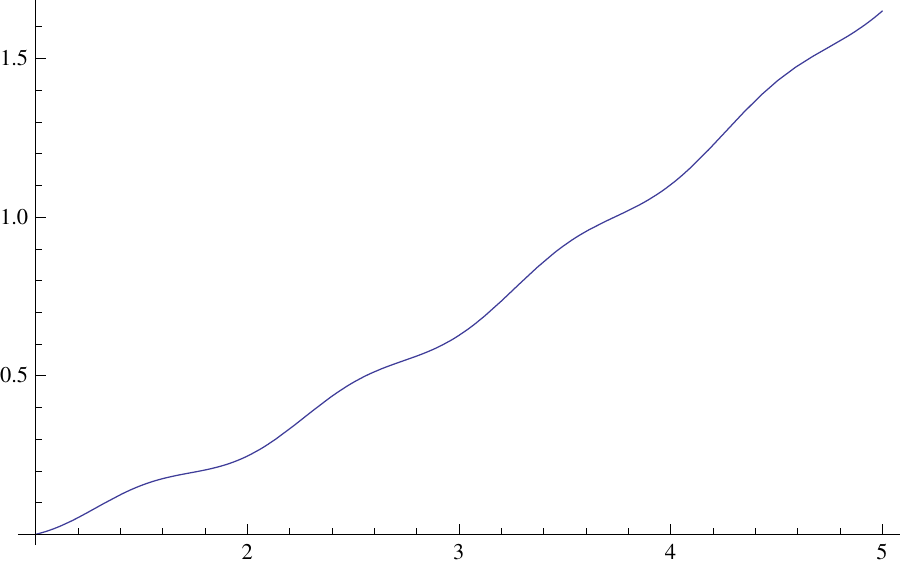}
\includegraphics[height=5cm]{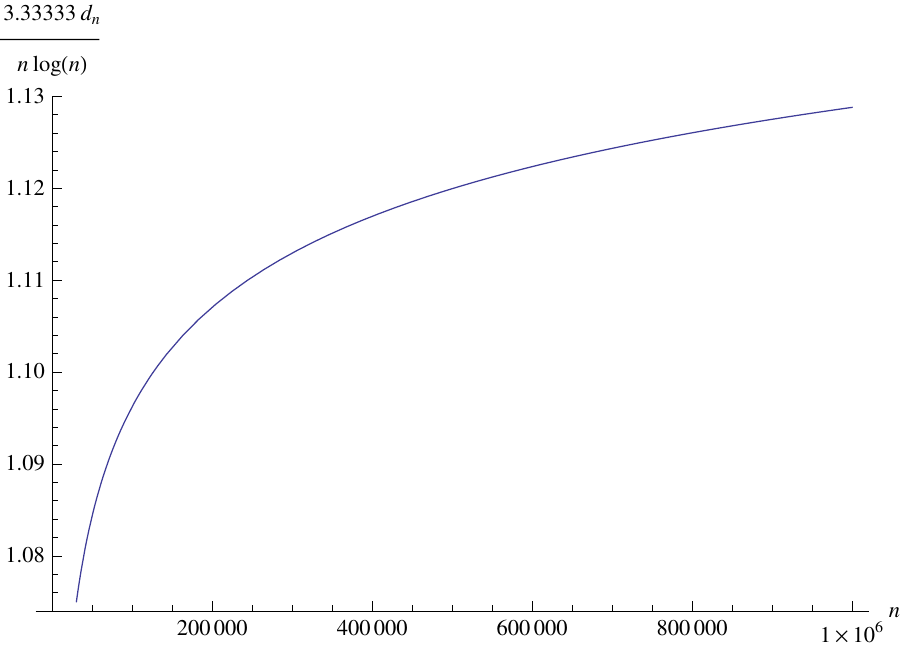}
\includegraphics[height=5cm]{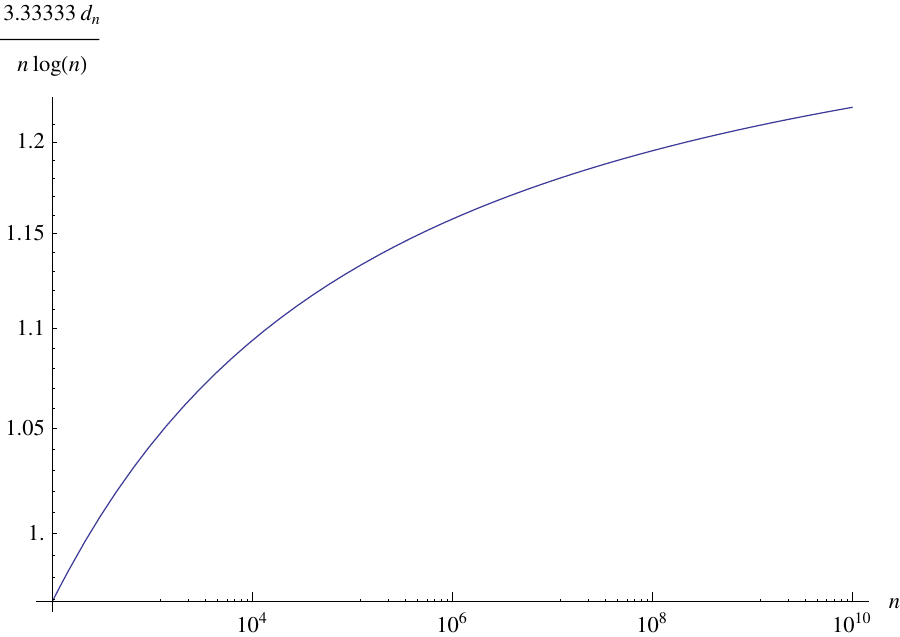}
\caption{  Anomalous exponents $d_{n}$ of  the moments of dissipation rate $\overline{{\cal E}^{n}}/\overline{{\cal E}}^{n}\propto Re^{d_{n}}$ given by formula (16).  From top to bottom: top:  $n\leq 5.0$,  middle:
$\frac{d_{n}}{0.3n\ln n}$  in the range $10^{5}\leq n\leq 10^{6}$;  bottom: the same in the broader interval  $10^{2}\leq n\leq 10^{10}$. }
\end{figure}

\noindent As  we  see from the Figure  2,  for  $n \gg 1$, the exponents $d_{n}\rightarrow 0.3 n\ln n$
 independent of magnitude of a constant $C$ meaning that at least in this limit the scaling exponents are universal.    
 
 
 

 
 



\noindent {\bf  Summary and Conclusions.}   1. This paper is based on equations (6)-(8) leading to   the well-defined Gaussian  low-order moments of velocity derivatives and increments.    In this case, ``Strong Turbulence `` is defined as a first appearance of  anomalous scaling exponents of  the moments of the dissipation rate. No  inertial range enters the considerations. \\
\noindent 2. In the spirit of Landau's  theory of transition,  it is assumed that in each statistical realization the transition occurs at  $\hat{R}_{\lambda,n}\geq 9.$,   independent on $n$.    The numerical value of transitional Reynolds number   $R_{\lambda}\approx 8.908$  was derived in Refs.[14]-[16].  Also, this result comes out  of  semi-empirical theories of large-scale turbulence modeling (Ref.[16]).\\ 
3.  {\bf  As follows from this paper, these two   very strong dynamic constraints  are  satisfied by formation of small-scale coherent structures  leading to   intermittency  (anomalous scaling) of dissipation fluctuations in turbulence.}\\
4.  The universality of these  results is  yet to be studied.  On one hand, recently  it has found some support  in numerical and experimental data on various flows [9],[17],[18]. On the other hand, as we see from Table.1,   numerical values of exponents $d_{n}$ may be  sensitive  to  statistics of the low-Reynolds number fluctuations.  \\  
 One argument in favor of broad universality can be found in Ref[11], (Model C),  where dynamic renormalization group was applied to the problem of the Navier-Stokes equations driven by various  random forces. The authors considered a general force (6) of an arbitrary statistics,  supported in a finite interval of wave-numbers $k\approx k_{f}$ and showed  that  in the limit $k<<k_{f}$ the velocity fluctuations, generated by the model,  obeyed Gaussian statistics. This result can be readily understood: each term of the  perturbation expansion of (6)-(8) is $O(k^{2n})$. Therefore, in the limit $k\rightarrow 0$, all high-order contributions with $n>1$ disappear as small. This situation corresponds to weak coupling.  It is not yet clear how universal this result is. \\
5.  The relation (16)  for  anomalous exponents $d_{n}$ is a consequence of  the coupling of  dissipation rate  and the {\bf random } fluctuations of  transitional Reynolds numbers (coupling constant)  studied in Ref.  [1].    The present paper is the first  where the role of  randomness of  a transition point  itself  in  the dynamics of small-scale   velocity fluctuations has been addressed.  It may be of interest to incorporate this feature  in the field-theoretical approaches, like Wyld's diagrammatic  expansions applied to the Navier-Stokes equations for small-scale fluctuations.  \\
6. In this paper  anomalous exponents of moments of velocity derivatives and those of dissipation rate have been calculated without introducing any adjustable parameters. 
 Given tremendous success of large-scale engineering simulations [16] , [19], which evolved into an indispensable design tool,  we tend to conclude that  Nelkin's question,`` In What Sense Turbulence is Unsolved Problem?'',  posed almost 25 years ago,  may need be discussed again.\\

 \begin{acknowledgements}     
\noindent V.Y. benefitted a lot from  detailed and illuminating  discussions of this work with  A.M.Polyakov. We are grateful to  H. Chen,  D. Ruelle, J. Schumacher,  I. Staroselsky, Ya.G. Sinai, K.R. Sreenivasan  and M.Vergassola    for many stimulating and informative discussions.  DD acknowledges support from NSF.
 \end{acknowledgements}          

\begin {references}
 \noindent  1. \  V.Yakhot \& D. Donzis,   ``Emergence of multi-scaling in a random-force stirred fluid'',  arXiv:1702.08468;  Phys.Rev.Lett. {\bf 119},  044501 (2017);\\
 \noindent 2. \ Kuz'min \& Patashinskii, ``Small Scale Chaos at Low Reynolds Numbers'', Preprint 91-20, Novosibirsk, (1991);\ Sov.Phys. JETP{\bf 49}, 1050 (1979);\\
\noindent 3.   \  C.Lissandrello,  K.L.Ekinci \& V.Yakhot, J. Fluid Mech, {\bf 778}, R3 (2015);\\
\noindent  4 . \  R.P.Feynman, ``The Feynman Lectures on Physics'', Addison Wesley Publishing, 1965.\\
\noindent  5. \     \ Monin and Yaglom, ``Statistical hydrodynamics'', MIT Press, 1975;  U. Frisch, ``Turbulence'', Cambridge University Press, 1995;l \\ 
\noindent 6. \  A.Polyakov, Sov.Phys.JETP, {\bf 32}, 296 (1971); ``Lectures Given at International  School on High Energy  Physics in Erevan",
23 November-4 December, 1971);\\
\noindent 7. \ M.Nelkin, Science {\bf 255}, 566 (1992);\\
\noindent 8. \ S.Y. Chen, B. Dhruva, S. K.R. Kurien, Sreenivasan \& M.A. Taylor 
`` 2005 Anomalous scaling of low-order
structure functions of turbulent velocity''  J. Fluid Mech. {\bf 533} 183 (2005); 
\ K.R. Sreenivasan  \& R.A.Antonia, Ann.Rev. of Fluid Mechanics {\bf 29} 435 (1997)\\
\noindent 9. \  J.Schumacher, K.R. Sreenivasan \& V. Yakhot, New J. of Phys.  {\bf 9}, 89 (2007); \\
\noindent 10. \  L.D.Landau \& E.M. Lifshitz, ``Fluid Mechanics'', Pergamon, New York, 1982; \\  
\noindent 11.  \ D. Forster, D. Nelson \&  M.J. Stephen, Phys.Rev.A {\bf 16}, 732 (1977);\\
 \noindent 12. \  H.W. Wyld, Ann.Phys.{\bf 14}, 143 (1961);\\
 \noindent 13. \  D.A. Donzis, P.K. Yeung \&  K.R. Sreenivasan,  
Phys.Fluids {\bf 20}, 045108 (2008); \\ 
 \noindent 14.  \ V. Yakhot \& L. Smith,  J. Sci.Comp. {\bf 7}, 35 (1992);\\
\noindent 15. \ V.Yakhot,, Phys.Rev.E, {\bf 90}, 043019 (2014);\\ 
\noindent 16 .  \  V. Yakhot, S.A. Orszag, T. Gatski, S. Thangam \& C.Speciale,   Phys. Fluids A{\bf 4}, 1510  (1992);   B.E.   Launder,and D.B. Spalding. Mathematical Models of Turbulence, Academic
Press, New York (1972);  B.E. Launder and D.B. Spaulding,  
Computer Methods in Applied Mechanics and engineering, {\bf 3}, 269 (1974).\\
\noindent  17. \   P.E. Hamlington, D. Krasnov, T. Boeck \&  J. Schumacher, 
J. Fluid. Mech. {\bf 701}, 419-429 (2012);\\ 
\noindent 18. \  J. Schumacher, J. D. Scheel, D. Krasnov, D. A. Donzis, V. Yakhot \&   K. R. Sreenivasan,
Proc. Natl. Acad. Sci. USA {\bf 111}, 10961-10965 (2014);\\
\noindent 19.  \ H. Chen et al.,   Science, {\bf 301} , 633 (2003). The annual sales of commercial packages  for numerical engineering simulations  are approaching $10^{9}$ USDollars. \\ 
 
\end{references}

\end{document}